\documentstyle[11pt,epsf]{article}

\def\lab#1        {\hbox{\small #1} }
\newcommand{\ben} {\begin{eqnarray*}}
\newcommand{\een} {\end{eqnarray*}}
\newcommand{\benn}{\begin{eqnarray}}
\newcommand{\eenn}{\end{eqnarray}}
\def\ineps        {\mathbin{\varepsilon}}

\def\be           {\begin{equation}}
\def\ee           {\end{equation}}
\def\Tr           {\mathop{\hbox{Tr}}}
\def\mb#1         {\mbox{\boldmath $#1$}}

\def\la           {\left\langle }
\def\ra           {\right\rangle }

\def\notesize     {\tiny }

\def\eff        {\hbox{\notesize eff}}
\def\diff       {\hbox{\notesize diff}}
\def\second     {\hbox{\notesize 2nd}}
\def\third      {\hbox{\notesize 3rd}}
\def\mon        {\hbox{\notesize mon}}
\def\const      {\hbox{const.}}

\begin{document}


\begin{tabbing}
\` Edinburgh 99/1 \\
\` OUTP-99-12P \\
\end{tabbing}
 
\vspace*{\fill}
\begin{center}
{\large\bf Monopole clusters, Z(2) vortices and confinement in SU(2).\\}
\vspace*{.25in}
{\large{\it UKQCD Collaboration}}\\
A. Hart$^1$
and M. Teper$^2$.\\
\vspace*{.2in}
{\small\sl $^1$Department of Physics, University of Edinburgh,
King's Buildings,\\ Edinburgh EH9 3JZ, Scotland.
\\
e--mail: hart@ph.ed.ac.uk
\\
\vspace{1ex}
$^2$Theoretical Physics, University of Oxford, 1 Keble Road,\\
Oxford OX1 3NP, England.
\\
e--mail: teper@thphys.ox.ac.uk}

\end{center}

\vfill
\begin{center}
\begin{minipage}{5in}
\begin{center}

{\bf Abstract.}
\end{center}

We extend our previous study of magnetic monopole currents 
in the maximally Abelian gauge
\cite{hart98}
to larger lattices at small lattice spacings
($20^4$ at $\beta = 2.5$ and $32^4$ at $\beta = 2.5115$). We confirm
that at these weak couplings there continues to be one monopole 
cluster that is very much longer than the rest and that the
string tension, $K$, is entirely due to it. The remaining 
clusters are compact objects whose population as a function of 
radius follows a power law that deviates from the scale 
invariant form, but much too weakly to suggest a link with the
analytically calculable size distribution of small instantons. 
We also search for traces of Z(2) vortices in the Abelian
projected fields; either as closed loops of `magnetic' flux
or through appropriate correlations amongst the monopoles.
We find, by direct calculation, that there is no confining condensate
of such flux loops. We also find, through the calculation of 
doubly charged Wilson loops within the monopole fields,
that there is no suppression of the $q=2$ effective
string tension out to at distances of at least  
$r \simeq 1.6/ \sqrt{K}$, suggesting that if there are
any vortices they are not encoded in the monopole fields.

\end{minipage}
\end{center}
\vfill
\noindent
PACS indices: 
11.15.Ha, 
12.38.Aw, 
14.80.Hv. 


\setcounter{page}{1}
\newpage
\pagestyle{plain}

\section{Introduction}
\label{sec_intro}

Many recent efforts to elucidate the mechanism of confinement in {\sc
qcd} and non--Abelian gauge theories have focused on isolating a
reduced set of variables that are responsible for the confining
behaviour. In the dual superconducting vacuum hypothesis
\cite{mandelstam76,thooft81}
the crucial degrees of freedom are the magnetic monopoles revealed
after Abelian projection. In the maximally Abelian gauge
\cite{thooft81,kronfeld87}
one finds that the extracted 
U(1) fields possess a string tension that approximately
equals the original SU(2) string tension (`Abelian dominance')
\cite{suzuki90},
and that this is almost entirely due to monopole currents in these 
Abelian fields (`monopole dominance')
\cite{stack94,bali96}.
The magnetic currents observed in the maximally Abelian gauge are 
found to have non-trivial correlations with gauge-invariant 
quantities such as the action and topological charge densities
(see for example
\cite{feurstein97,bakker98}
and references therein)
and this invites the hypothesis that the structures formed by the 
magnetic monopoles correspond to similar objects in the SU(2) vacuum,
seen after gauge fixing and Abelian projection.  If the magnetic
monopoles truly reflect the otherwise unknown infrared physics of the
SU(2) vacuum, analysis of these structures may provide important
information about the confinement mechanism.

The main purpose of this paper is to extend our previous study
\cite{hart98}
of monopole currents to lattices that are larger in physical units 
at the smallest lattice spacings. As reviewed in Sec.~\ref{sec_cl_str},
we obtained in
\cite{hart98}
a strikingly simple monopole picture at $\beta = 2.3$,~2.4. When
the magnetic monopole currents are organised into separate clusters, 
one finds in each field configuration one and only one cluster 
which is very much larger than the rest and which percolates 
throughout the entire lattice volume.  Moreover this largest
cluster is alone responsible for infrared physics such as the string
tension. The remaining clusters are compact objects with radii varying
with length roughly as $r \propto \sqrt{l}$ and with a population
that follows a power law as a function of length. We found the
exponent of this power law to be consistent with a universal
value of 3. This simple pattern became more confused at 
$\beta = 2.5$. The scaling relations for cluster size
that we established in
\cite{hart98}
suggested that our $L=16$ lattice at $\beta = 2.5$ was simply too
small. There was of course an alternative possibility: that the simple
picture we found at lower $\beta$ failed as one approached 
the continuum limit. Clearly it is important to distinguish between
these two possibilities, and this is what we propose to do in this
paper. The cluster size scaling relations referred to above
imply that an $L=32$ lattice at $\beta=2.5115$ should have a
large enough volume. Such gauge fixed lattice fields were
made available to us by G. Bali and we have used them, supplemented
by calculations on an intermediate $L=20$ volume at $\beta=2.5$, 
to obtain evidence, as described in Secs.~\ref{sec_cl_str} and
~\ref{sec_mon_vor_K}, that the monopole picture we found previously 
is in fact valid at these lattice spacings and that the
deviations we found previously were due to too small
a lattice size.

The fact that one has to go to space-time volumes that are ever 
larger, in physical units, as the lattice spacing decreases,
hints at some kind of breakdown of `monopole dominance'
in the continuum limit. We finish Sec.~\ref{sec_cl_str},
with a discussion of the form that this breakdown might
take.

An attractive alternative to the dual superconducting vacuum as
a mechanism for confinement is vortex condensation
\cite{thooft79,mack80,nielsen79,kovacs98,deldebbio98,ambjorn98}.
Here the confining degrees of freedom are the vortices created
by the 't Hooft dual disorder loops
\cite{thooft79}
and the confining disorder is located in the 
centre Z(N) of the SU(N) gauge group. When such a vortex
intertwines a Wilson loop, the fields along the loop undergo a 
gauge transformation that varies from unity to a non-trivial 
element of the centre as one goes once around the Wilson loop.
For SU(2) this means that the Wilson loop acquires
a factor of $-1$. A condensate of such vortices will therefore
completely disorder the Wilson loop and will lead to linear
confinement. At the centre of the vortex, which will be a line
in $D=2+1$ and a sheet in $D=3+1$, the fields are clearly
singular (multivalued) if we demand that the vortex correspond
to a gauge transformation almost everywhere. In a properly 
regularised and renormalised theory, this singularity will
be smoothed out 
\cite{thooft79}
into a core of finite size in which there is a non-trivial but
finite action density, and whose size will be $O(1)$
in units of the physical length scale of the theory.
One can either try to study these vortices
directly in the SU(2) gauge fields or one can go to the centre gauge
\cite{deldebbio98,ambjorn98},
where one makes the gauge links as close to $+1$ or $-1$ as possible,
and construct the corresponding fields where the link matrices take 
values in Z(2) (`centre projection') and where the only 
nontrivial fluctuations are singular Z(2) vortices. 
Just as a 't Hooft--Polyakov monopole  will appear
as a singular Dirac monopole in the Abelian fields that one obtains
after Abelian projection, one would expect the presence of
a vortex in the SU(2) fields to appear as a singular Z(2) vortex
after centre projection. This picture has received increasing attention
recently and has, for example, proved successful in reproducing
the static quark potential
\cite{kovacs98,deldebbio98}
(`centre dominance'). 
Our ability, in this paper, to address the question of how 
important are such vortices is constrained by the 
fact that we only work with Abelian projected SU(2) fields. 
So first we need to clarify how such vortices might be encoded
in these Abelian fields and only then can we perform numerical
tests to see whether there is any sign of their presence.
This is the content of Sec.~\ref{sec_mon_vor_K}.

Finally there is a summary of the results in Sec.~\ref{sec_summ}.

\section{Monopole cluster structure}
\label{sec_cl_str}

\subsection{Background}

Fixing to the maximally Abelian gauge of SU(2) amounts to maximising
with respect to gauge transformations the Morse functional
\be
R = - \sum_{n,\mu} \Tr \left( U_\mu(n) \cdot i\sigma_3 \cdot 
U^\dagger_\mu(n) \cdot i\sigma_3 \right).
\label{eqn_R}
\ee
It is easy to see that this maximises the matrix elements
$|[U_\mu(n)]_{11}|^2$ summed over all links. That is to say, it is the
gauge in which the SU(2) link matrices are made to look as diagonal,
and as Abelian, as possible --- hence the name. Having fixed to this
gauge, the link matrices are then written in a factored form and the
U(1) link angle (just the phase of $[U_\mu(n)]_{11}$) is extracted.
The U(1) field contains integer valued monopole currents
\cite{degrand80},
$\{ j_\mu(n) \}$,
which satisfy a continuity relation, 
$\Delta_\mu j_\mu(n) = 0$,
and may be unambiguously assigned to one of
a set of mutually disconnected closed networks, or `clusters.'

In
\cite{hart98}
we found that the clusters may be divided into two classes on 
the basis of their length, where the length is obtained by simply 
summing the current in the cluster
\be
l = \sum_{n,\mu \ineps \mbox{\small cl.}}
\left| j_\mu(n) \right|.
\ee
The first class comprises the largest cluster, which is physically the
most interesting. It percolates the whole lattice volume and its
length $l_{\max}$ is simply proportional to the volume $L^4$ 
(at least in the interval
$2.3 \le \beta \le 2.5$) 
when these are re-expressed in physical units, {\it i.e.}
$l_{\max}\sqrt{K} \propto (L\sqrt{K})^4$,
where $K$ is the SU(2) lattice string tension in lattice
units and we use $1/\sqrt{K}$
to set our physical length scale.  We remark that over this range in
$\beta$ there is a factor 2 change in the lattice spacing, and so
one might have expected that the extra  ultraviolet fluctuations on 
the finer lattice would lead to significant violations of the 
na\"{\i}ve scaling relation. That is to say, one might expect to
need to coarse grain the currents at larger $\beta$ to obtain 
reasonable scaling. That this is not required is perhaps surprising.

The remaining clusters were found to be much shorter. Their population
as a function of length (the `length spectrum') is described by a 
power law
\be
N(l) = \frac{c_l(\beta)}{l^\gamma},
\label{eqn_len_spec}
\ee
where $\gamma \approx 3$ for all lattice spacings and sizes
tested and the coefficient $c_l(\beta)$ is proportional to the 
lattice volume, $L^4$, and depends weakly on $\beta$.
The radius of gyration of these clusters is small and
approximately proportional to the square root of the cluster
length, just like a random walk. When folded with the length 
spectrum, this suggests 
\cite{hart98}
that the `radius spectrum' should also be described by a power law
\be
N(r) = \frac{c_r(\beta)}{r^\eta},
\label{eqn_rad_spec}
\ee
with $\eta \approx 5$ and $c_r(\beta)$ weakly dependent on
$\beta$. Such a spectrum is close to the scale invariant spectrum of
4--dimensional balls of radius $\rho$, $N(\rho) d\rho \sim d\rho/\rho
\times 1/\rho^4$, and so one might try and relate these clusters to
the SU(2) instantons in the theory, which classically also possess a 
scale-invariant spectrum. It is well known, however, that the inclusion
of quantum corrections renders the spectrum of the latter far from 
scale invariant, at least for the small instantons where perturbation
theory can be trusted, and so such a
connexion does not seem to be possible
\cite{hart98}.

On sufficiently large volumes the difference in length between the
largest and second largest cluster is very marked, and where this gulf
is clear one finds that the long range physics such as the monopole 
string tension arises solely from the largest cluster. This is the 
case at $\beta = 2.3$, $L \ge 10$ and at  $\beta = 2.4$, 
$L \ge 16$. On moving to a finer $L=16$ lattice at $\beta=2.5$ 
the gulf was found to disappear and the origin of the long range 
physics was no longer so clear cut. This could be a mere finite
volume effect, or, much more seriously, it might signal the
breakdown of this monopole picture in the weak coupling, continuum
limit. Clearly this needs to be resolved and the only unambiguous
way to do so is by performing the calculations on large enough lattices.

\subsection{This calculation}
\label{ssec_thiscalc}

The direct way to estimate the lattice size necessary at $\beta=2.5$ 
to restore (if that is possible) our picture is as follows.
Suppose that the average size of the  second largest cluster
scales approximately as
\be
l_{\second} \propto L^\alpha (\sqrt{K})^\delta .
\ee
We know that $l_{\max} \propto L^4 (\sqrt{K})^3$ to a good
approximation for the largest cluster. So we will maintain the same
ratio of lengths $l_{\second} / l_{\max}$, and a gulf between these,
if
\be
\frac{L_1}{L_2} = \left( \frac{\sqrt{K_1}}{\sqrt{K_2}} \right)^{
-\left(\frac{3-\delta}{4-\alpha} \right)}
\label{eqn_lK_scale}
\ee

If we take our directly calculated values of $l_{\second}$, they seem
to give roughly $\alpha \simeq 1$ and $\delta \simeq -2$.  This
suggests that we need to scale our lattice size with $\beta$ so as to
keep $L(\sqrt{K})^{5/3}$ constant. This estimate is not entirely
reliable because, on smaller lattices, the distributions of the
`largest' and `second largest' clusters overlap so that they exchange
{\it r\^{o}les}. An alternative estimate can be obtained from the tail
of the distribution in eqn.~\ref{eqn_len_spec} that integrates to
unity. Doing so
\cite{hart98}
one obtains $\alpha \simeq 2$ and $0 < \delta < 0.25 $.
This suggests that we scale our lattice size so as to keep 
$L(\sqrt{K})^{\{1.4 \to 1.5\}}$ constant. This estimate
is also not very reliable, since it assumes that the
distribution of secondary cluster sizes on different
field configurations fluctuates no more than mildly about
the average distribution given in eqn.~\ref{eqn_len_spec}.
In fact the fluctuations are very large. [As we can see
immediately when we try to calculate $\langle l^2 \rangle$
in order to obtain a standard fluctuation --- it diverges
for a length spectrum with $N(l)\propto dl/l^3$.] Nonetheless, the 
two very different estimates we have given above produce
a very similar final criterion: to maintain the same
gap between the largest and second largest clusters
as $\beta$ is varied, one should choose  $L$ so as to
keep $L(\sqrt{K})^{\sim 1.5}$ constant.

So if we wish to match the clear picture on an $L=10$ lattice at
$\beta=2.3$ (where $K = 0.136$~(2)) we should work on a lattice that
is roughly $L=28$ at $\beta=2.5$ (where $K = 0.0346$~(8)). In
particular we note that an $L=32$ lattice at $\beta = 2.5115$ (where
$K = 0.0324$~(10)) is more than large enough and an ensemble of 100
such configurations, already gauge fixed
\cite{bali96},
has been made available to us by the authors. The gauge fixing 
procedure used in obtaining these is somewhat different
from the one we have used in our previous calculations 
(in its treatment of the Gribov copies --- see below)
and although this is not expected to affect the qualitative features
that are our primary interest here, it will have some effect on
detailed questions of scaling etc. We have therefore also
performed a calculation on an ensemble of 100 gauge fixed $L=20$
field configurations at $\beta=2.5$. While the latter volume
is not expected to be large enough to recreate a clear gulf
between the largest and remaining clusters, we would expect to find 
smaller finite size corrections than with the $L=16$ lattice
we used previously.

In gauge fixing a configuration we select a local maximum of the Morse
functional, $R$, of which on lattices large enough to support
non--perturbative physics there are typically a very large number
\cite{hart97a}.
These correspond to the (lattice) Gribov copies. Gauge dependent
quantities appear to vary by ${\cal O}(10\%)$ depending upon the
Gribov copy chosen; this is true not only of local quantities such as
the magnetic current density
\cite{hioki91}
but also of supposedly long range, physical numbers such as the 
Abelian and monopole string tensions
\cite{bali96,hart97a}.
Some criterion must be employed for the selection of the maxima of
$R$, and in the absence of a clear understanding of which maximum, if
any, is the most `physical', one maximum was selected at random in
\cite{hart98}. 
An alternative strategy, used in gauge fixing the $L=32$ lattices at
$\beta = 2.5115$, is to pursue the global maximum of $R$
\cite{bali96}.
Each field configuration is fixed to the maximally Abelian gauge
10~times using a simulated annealing algorithm that already weights
the distribution of maxima so selected towards those of higher
$R$. The solution with the largest $R$ from these is selected. Details
of this method are discussed in
\cite{bali96}.
The difference in procedures invites caution in comparing exact numbers
between this ensemble and those studied previously; for example a
${\cal O}(10\%)$ suppression in the string tension is observed. 
It is likely that cluster lengths will differ by a corresponding
amount and this will prevent a quantitative scaling analysis
using this ensemble. The power law indices do appear, however, to be robust
\cite{hart97b}
and it also seems likely that ratios of string tensions obtained on the
same ensemble can be reliably compared with other ratios.

\subsection{Cluster properties}

The fact that the largest cluster does not belong to the same
distribution as the smaller clusters is seen from the very 
different scaling properties of these clusters with volume
\cite{hart98}.
It is also apparent from the fact that the largest cluster is very
much longer than the second largest cluster. Indeed for a large enough
volume and for a reasonable size of the configuration ensemble, there
will be a substantial gulf between the distribution of largest cluster
lengths and that of the second largest clusters. By contrast the
length distributions of the second and third largest clusters strongly
overlap. This is the situation that prevailed for the larger lattices
at $\beta=2.3$ and 2.4 but which broke down on the $L=16$ lattice at
$\beta=2.5$. We can now compare what we find on our $L=20$ and $L=32$
lattices with the latter.  This is done in
Table~\ref{table_length}. There we show the longest and shortest
cluster lengths for the largest, second largest and third largest
cluster respectively over the ensemble.  The ensemble sizes are not
exactly the same, but it is nonetheless clear that there is a real
gulf between the largest and second largest clusters on the $L=32$
lattice while there is significant overlap in the $L=16$ case. The
$L=20$ lattice is a marginal case. We conclude from this that the
apparent loss of a well separated largest cluster as seen in
\cite{hart98}
at $\beta=2.5$ was in fact a finite volume effect, and that
our scaling analysis has proved reliable in predicting what 
volume one needs to use in order to regain the simple picture.

\begin{figure}[p]
\begin{center}
\leavevmode
\hbox{
\epsfxsize = 4in
\epsffile{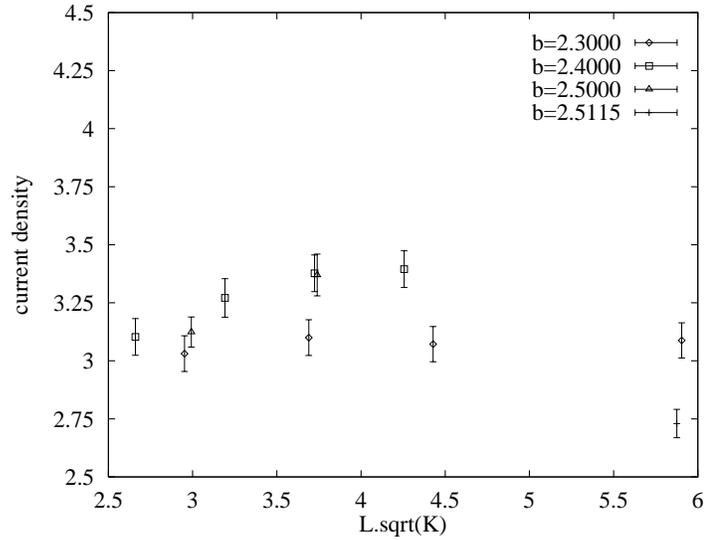}
}
\end{center}
\caption{The current density of the largest cluster as a
function of lattice size in physical units for various $\beta$.}
\label{fig_curr_dens}
\end{figure}

In Figure~\ref{fig_curr_dens} we show how the length of the largest
cluster varies with the lattice volume when both are expressed in
physical units (set by $\sqrt{K}$). To be specific, we have divided
$l_{\max}\sqrt{K}$ by $(L\sqrt{K})^4$ and plotted the resulting
numbers against $L\sqrt{K}$ for both our new and our old
calculations. The fact that at fixed $\beta$ the values fall on a
horizontal line tells us that that the length of the largest cluster
is proportional to the volume at fixed lattice spacing: $l_{\max}
\propto L^4$. The fact that the various horizontal lines almost
coincide tells us that the current density in the largest cluster is
consistent with scaling. That is to say, it has a finite non-zero
value in the continuum limit. Thus the monopole whose world line
traces out this largest cluster, percolates throughout the space--time
volume and its world line is sufficiently smooth on short distance
scales that its length does not show any sign of diverging as we take
the continuum limit. We note that the $L=32$ lattice deviates by $\sim
10\%$ from the other values. This is consistent with what we might
have expected from the different gauge fixing procedure used in that
case.

\begin{figure}[p]
\begin{center}
\leavevmode
\hbox{
\epsfxsize = 3in
\epsffile{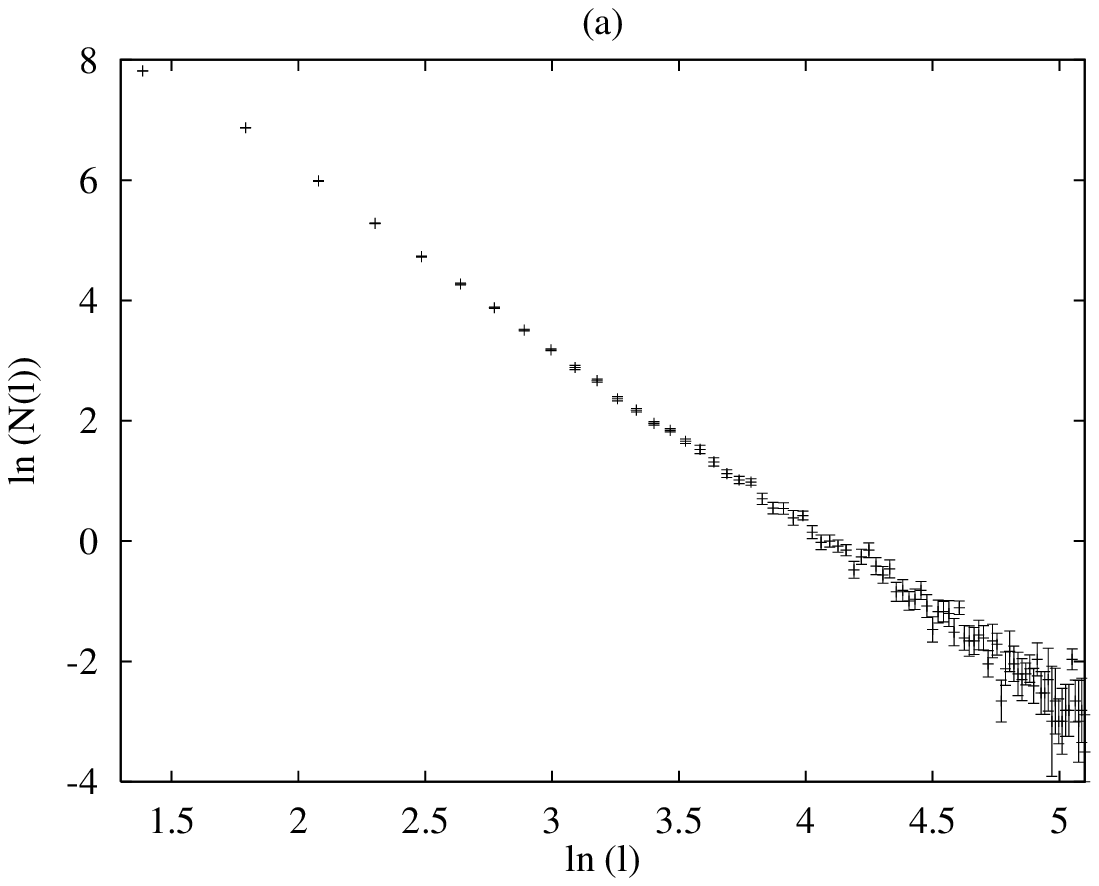}
\epsfxsize = 3in
\epsffile{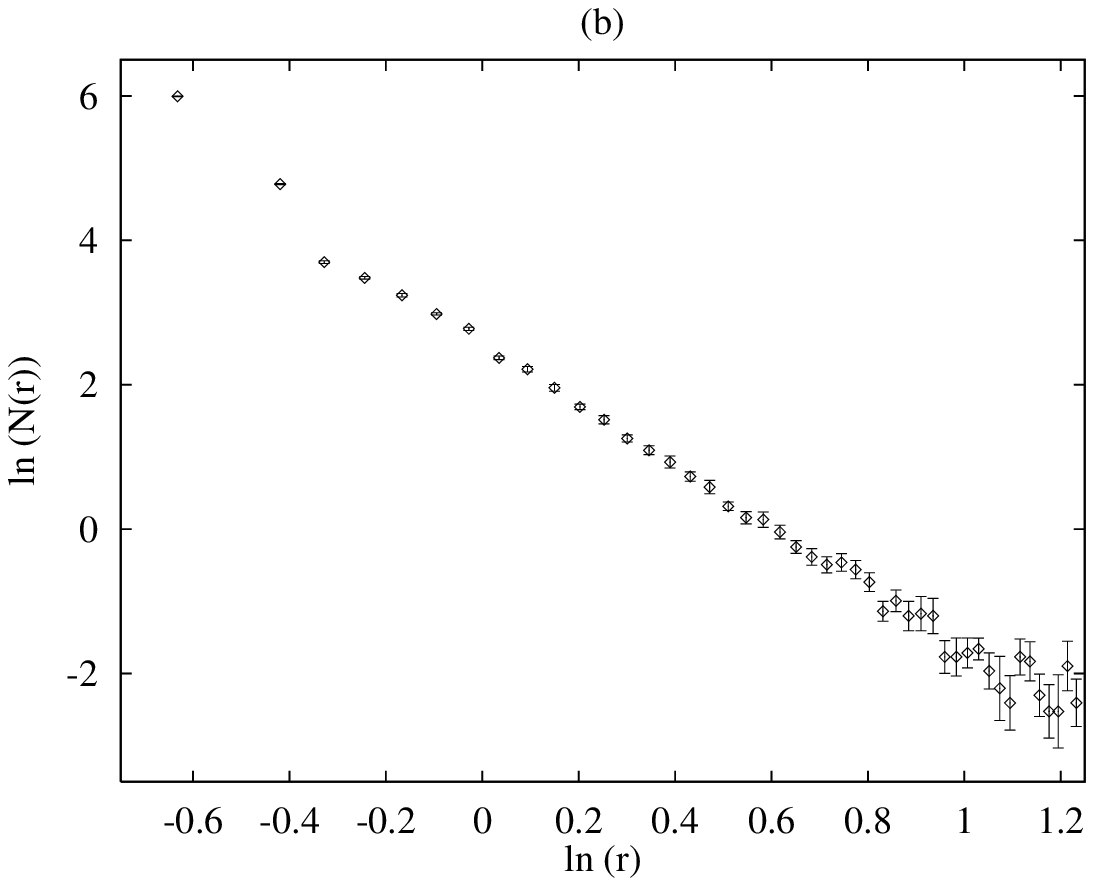}
}
\end{center}
\caption{The cluster spectra by a) length and b) radius at $\beta = 2.5115$
on $L = 32$.}
\label{fig_len_spec}
\end{figure}

Turning now to the secondary clusters, we display in
Figure~\ref{fig_len_spec} the length spectrum that we obtain at
$\beta=2.5115$.  It is clearly well described by a power law as in
eqn.~\ref{eqn_len_spec} and we fit the exponent to be $\gamma =
3.01$~(8). This is in accord with the universal value of 3 that was
postulated in
\cite{hart98}
on the basis of calculations on coarser lattices. The
value one fits to the spectrum obtained on the $L=20$
lattice at $\beta=2.5$ is $\gamma = 2.98$~(7) and
is equally consistent. We also examine the dependence
on $\beta$ of the coefficient $c_l(\beta)$ in 
eqn.~\ref{eqn_len_spec} adding to the older work our calculations 
at $\beta=2.5$ on the $L=20$ lattice. [We do not use
the $L=32$ lattice for this purpose because of
the different gauge fixing procedure used.]
If we assume a constant power (which is approximately the
case), then  $c_l(\beta)$ is just proportional to the total length 
of the secondary clusters. At fixed $\beta$ we find this length
to be proportional to $L^4$ just as one might expect. [Small
clusters in very different parts of a large volume are
presumably independent.] The dependence on $\beta$, on
the other hand, is much less clear. Between $\beta=2.3$
and $\beta=2.4$ it varies weakly, roughly as $K^{0.12 \pm 0.13}$.
Between $\beta=2.4$ and $\beta=2.5$ it varies more strongly, 
roughly as $K^{0.48 \pm 0.09}$. We can try to summarise this 
by saying that
\be
c_l(\beta) = \const L^4 {\sqrt{K}}^{\zeta}
\label{eqn_coeff_spec}
\ee
where $\zeta = 0.5 \pm 0.5$, which is consistent with what was found
previously
\cite{hart98}.

The smaller clusters are compact objects in $d=4$, and having
determined the cluster spectrum as a function of length we can then
ask what is the spectrum when re-expressed as a function of the radius
(of gyration) of the cluster.  In
\cite{hart98}
we obtained this spectrum by determining the average radius as a
function of length, and folding that in with the number density as a
function of length. This is an approximate procedure (forced upon us
by the fact that we did not foresee the interest of this spectrum
during the processing of the clusters) and one can obtain the spectrum
more accurately by calculating $r$ for each cluster and forming the
spectrum directly. Doing so for the $L = 32$ lattice at $\beta =
2.5115$, also in Figure~\ref{fig_len_spec}, we find a power law as in
eqn.~\ref{eqn_rad_spec} with $\eta = 4.20$~(8). The spectrum on the
$L=20$ lattice at $\beta=2.5$ yields $\eta = 4.27$~(6).  We recall
that in
\cite{hart98}
we claimed that the spectrum was consistent with the scale invariant
result $dr/r \times 1/r^4$, {\it i.e.} $\eta = 5$. This followed from
the fact that we found the the radius of the smaller clusters to vary
with their length as $r(l) = s + t.l^{0.5}$ {\it i.e.} just what one
would expect from a random walk. Folded with a length spectrum $N(l)
\sim 1/l^3$, this gives $\eta = 5$. On the $L=32$ lattice we still
find that the random walk {\it ansatz} provides an acceptable fit but
we also find that $r(l) = s + t.l^{0.65}$ works equally well over
similar ranges. The latter, when folded with $\gamma = 3$, gives $\eta
= 4.2$. It is clear that the direct calculation of $N(r)$ is much more
accurate than the indirect approach.

Treating the power as a free parameter in the fit, $r(l) = s + t.l^u$,
we find $u = 0.57$~(3) on $L=32$ at $\beta = 2.5115$, consistent with
$u = 0.58$~(4) on $L=20$ at $\beta=2.5$. Thus both $u=0.5$ and
$u=0.65$ lie within about two standard deviations from the fitted value. 
Note that what the fitted powers $\gamma$ and $u$ parameterise 
are the means of the distributions of lengths and radii 
respectively. That combining these does not give the 
directly calculated value of $\eta$ is not unexpected, and
reflects the importance of fluctuations around the mean in the
distributions.

If the secondary monopole clusters can be associated with localised
excitations of the full SU(2) vacuum (`4--balls'), it would seem that
such objects do not have an exactly scale invariant distribution in
space--time, so that the number of larger radius objects is somewhat 
greater than would be expected were this the case. Now it is known
that an isolated instanton (even with quantum fluctuations) 
is associated with a monopole cluster within its core
(see
\cite{hart96,inst99}
and references therein)
and that the scale invariant semiclassical density of instantons
acquires corrections due to quantum fluctuations. These corrections
are, however, very large; in SU(2) the spectrum of small instantons
(where perturbation theory is reliable) goes as $N(\rho) d\rho \propto
d\rho/\rho \times \rho^{10/3}$, where $\rho$ is the core size. The
scale breaking we have observed for monopole clusters is negligible in
comparison. Thus we cannot identify the `4--balls' with
instantons. Indeed, the fact that the monopole spectrum is so close to
being scale invariant strongly suggests that these secondary clusters
have no physical significance. In the next section we shall show
explicitly that, in the large volume limit, they do not play any part
in the long range confining physics.

\subsection{Breakdown of `monopole dominance'?}

We finish this section by asking if there are hints from our cluster
analysis that `monopole dominance' might be breaking down as we
approach the continuum limit. This question is motivated by the
observation that the monopoles are identified by a gauge fixing
procedure which involves making the bare SU(2) fields as diagonal as
possible. Since the theory is renormalisable, the long distance
physics increasingly decouples from the fluctuations of the
ultraviolet bare fields as we approach the continuum limit. For
example, the ultraviolet contribution to the action density is
$O(1/\beta)$ while the long distance contribution is
$O(e^{-c\beta})$. Thus as $a\to 0$ the maximally Abelian gauge will be
overwhelmingly driven by ultraviolet rather than by physical
fluctuations.  Moreover at the location of the monopoles the Abelian
fields are far from unity and so one would expect the SU(2) fields
also to be far from unity. Thus the number of monopoles would seem to
be constrained by the probability of finding corresponding clumps of
SU(2) fields with large plaquette values. This probability depends on
the detailed form of the SU(2) lattice action far from the Gaussian
minimum and one could easily choose an action where it is completely
suppressed and yet which one would expect to be in the usual
universality class.  None of the above arguments are completely 
compelling of course. In the Gaussian approximation, for example, 
the  $O(1/\beta)$ ultraviolet fluctuations would not generate any
monopoles at all, and in that case there would be no reason to
expect any breakdown of monopole dominance. Nonetheless 
the arguments do suggest that it would be surprising if the long
distance physics were to be usefully and simply encoded in the 
monopole structure (as defined on the smallest ultraviolet scales) 
all the way to the continuum limit.

There are different ways in which monopole dominance
could be lost. The most extreme possibility is that 
as $a \to 0$ the fields simply cease to contain 
monopole clusters that are large enough to disorder
large Wilson loops. That this is indeed so has been argued in
\cite{grady98}
where it has been claimed that the exponent $\gamma$
in our eqn.~\ref{eqn_len_spec} (but defined for loops 
rather than for clusters) increases rapidly with 
decreasing $a$. Of course this would 
not in itself preclude the existence of a large percolating 
cluster, as long as this cluster could be decomposed
into a large number of small and correlated intersecting loops.
Irrespective of this, we also note that the volumes used in
\cite{grady98}
are very small by the criterion given in eqn.~\ref{eqn_lK_scale}. For
example, from our scaling relations we would expect to need an $L
\simeq 46$ lattice at $\beta=2.6$ and an $L \simeq 70$ lattice at
$\beta=2.7$ in order to resolve our simple monopole picture, if it
still holds at these values of $\beta$. This contrasts with the
$L=12$ and $L=20$ lattices actually employed in
\cite{grady98}.
So it appears to us that while the claims in
\cite{grady98}
are certainly interesting, further calculations
on much larger lattices are required.
 
Our work suggests a somewhat different form of the breakdown
to the one above. We see 
from eqn.~\ref{eqn_coeff_spec} that the ratio of the (total)
monopole current residing in the physically irrelevant, smaller 
clusters to that residing in the large percolating cluster, 
increases rapidly as  $a\to 0$ as 
$1/\sqrt{K}^{3-\zeta} \propto 1/a^{3-\zeta}$.
This suggests that as $a\to 0$ a calculation
of Wilson loops will become increasingly dominated 
by the fluctuating contribution of the unphysical
monopoles that are  ever denser on physical length scales,
and that this will eventually prevent us from extracting
a potential or string tension. That is to say: 
calculations in the maximally Abelian gauge will eventually
acquire a similar problem to that which typically afflicts Abelian 
projections using other gauges. In our case we can overcome
this problem by going to a large enough volume that the
physically relevant percolating cluster can be simply 
identified. [The reason this cannot be done with other
typical Abelian gauge fixings is that there the unphysical
monopoles are dense on lattice scales making any meaningful
separation into clusters impossible.] We can then extract the
string tension using, in our Wilson loop calculation, only
this largest monopole cluster. The fact that the length of
this cluster scales in physical units, with apparently
no significant anomalous dimension, tells us that this
calculation will not be drowned in ultraviolet `noise' as we 
approach the continuum limit. Of course, the fact that we can 
only do this for volumes that diverge in physical units as 
$a\to 0$ is a symptom of the underlying breakdown of the
Abelian projection.

The qualitative discussion in the previous paragraph over-estimates
the effect of the secondary clusters; for example, the contribution
that a cluster of fixed size in lattice units makes to a Wilson loop
of a fixed physical size will clearly go to zero as $a\to 0$. So it is
useful to ask how Wilson loops are affected by the secondary clusters,
and to do so using approximations that underestimate the effect of
these smaller clusters.  Consider an $R\times R$ Wilson loop.  A
monopole cluster that has an extent $r$ that is smaller then $R$ will
affect it only weakly through higher multipole fields which cannot on
their own give rise to an area law decay and a string tension. So we
neglect such clusters and consider only those larger than $R$.  Let us
first neglect the observed breaking of scale invariance and simply
assume that $r \propto \sqrt{l}$ and that $\gamma = 3$.  We then find,
by integrating eqn.~\ref{eqn_len_spec} and using
eqn.~\ref{eqn_coeff_spec}, that the number of secondary clusters with
$r>R$ is proportional to $L^4 \sqrt{K}^{\zeta}/R^4$. We further assume
that such clusters must be within a distance $\xi$ from the minimal
surface of the Wilson loop, where $\xi$ is the screening length, if
they are to disorder that loop significantly.  The lattice volume this
encompasses is the area of the planar loop, $R^2$, multiplied by a
factor of $\xi$ for each of the two orthogonal directions in $d=4$.
So the probability for this Wilson loop to be disordered thus
decreases with $R$ as $(R^2\xi^2/L^4 \times L^4
\sqrt{K}^{\zeta}/R^4)
\sim \sqrt{K}^{\zeta} (\xi/R)^2$. So if we look
at a Wilson loop that is of a fixed size in physical units, {\it i.e.}
$R / \xi$ fixed assuming $\xi$ scales as a physical quantity
\cite{hart98}, 
then the influence of the secondary clusters will decrease to zero as
$a\to 0$ as long as $\zeta > 0$. If $\zeta < 0$, however, then we
would have to go to Wilson loops that were ever larger in physical
units as we approached the continuum limit, in order that the physical
contribution from the percolating cluster should not be swamped by the
unphysical contribution of the secondary clusters. Of course this
calculation uses a scale invariant $dr/r \times 1/r^4$ spectrum,
whereas, as we have seen, there is significant scale breaking
and the actual spectrum is closer to $dr/r
\times 1/r^{3.2}$.  If we redo the above analysis with the latter
spectrum we see that we are only guaranteed to preserve this aspect of
`monopole dominance' if $\zeta >0.8$.  As demonstrated in
eqn.~\ref{eqn_coeff_spec}, there is some evidence that $\zeta>0$ but
it is not at all clear that $\zeta >0.8$. All this indicates that even
in a calculation that errs on the side of neglecting the effect
of the smaller clusters, they nonetheless will most likely dominate  
the values of Wilson loops on fixed physical length scales.
It is only by separating the percolating cluster from the
other smaller clusters, and calculating Wilson loops just using
that largest cluster, that we can hope to be able to
extract the string tension as $a\to 0$.

\section{Monopoles, vortices and the string tension}
\label{sec_mon_vor_K}

In this section we begin by describing how we calculate the string 
tension from an arbitrary set of monopole currents. We then go on
to show that even at the smallest lattice spacings, the string
tension arises essentially entirely from the largest cluster,
as long as we use a sufficiently large volume.
We then calculate the string tension for sources that have
a charge of $q=2$,~3 and~4 times the basic charge, and compare 
these results to a simple toy model calculation. Finally
we discuss the implications of our calculations for the 
question whether it is really monopoles or vortices that
drive the confining physics.

\subsection{Monopole Wilson loops}
\label{ssec_mon_wl}

The monopole contribution to the string tension may be estimated using
Wilson loops. If the magnetic flux due to the monopole currents
through a surface spanning the Wilson loop, $\cal C$, (by default the
minimal one) is $\Phi({\cal S})$, then the charge $q$ Wilson loop has
value
\be
W({\cal C}) = \exp [ iq\Phi({\cal S}) ].
\ee
We may obtain the static potential from the rectangular Wilson loops 
\be
V(r) = \lim_{t \to \infty} V_{\eff}(r,t) \equiv
\lim_{t \to \infty} \ln \left[
\frac{\la W(r,t) \ra}{\la W(r,t+a) \ra} \right].
\ee
The string tension, $K$, may then be obtained from the long
range behaviour of this potential, $V(r) \simeq Kr$. The string
tension may also be found from the Creutz ratios
\be
K = \lim_{r \to \infty} K_{\eff}(r) \equiv \lim_{r \to \infty}
\ln \left[
\frac{\la W(r+a,r) \ra \la W(r,r+a) \ra}
{\la W(r,r) \ra \la W(r+a,r+a) \ra} 
\right].
\ee
Square Creutz ratios at a given $r$ are useful because they 
provide a relatively precise probe for the existence of
confining physics on that length scale. In addition Creutz
ratios are useful where the quality of the `data' precludes 
the double limit of the potential fit. This is so particularly
when positivity is badly broken as it frequently is for our
gauge dependent correlators.

The magnetic flux due to the monopole currents is found by solving a
set of Maxwell equations with a dual vector potential reflecting the
exclusively magnetic source terms. An iterative algorithm being
prohibitively slow on $L=32$, we utilised a fast Fourier transform method
to evaluate an approximate solution as the convolution of the periodic
lattice Coulomb propagator and the magnetic current sources
\cite{stack92}. 
The error in this solution was then reduced to an
acceptable level by using it as the starting point 
for the over--relaxed, iterative method.

We may use any subset of the monopole currents as the source term
to calculate the contribution to the Wilson loops and potential of
those currents, provided that they i) are locally conserved and ii)
have net zero winding number around the periodic lattice in all
directions, {\it e.g.}
\be
Q_{\mu=4} \equiv \sum_{x,y,z} j_4(x,y,z,t=1) = 0.
\ee
If we choose complete clusters, then the first condition is always
satisfied but the second condition is often not met (even though the
winding number for all the clusters together must be zero). In such
cases we introduce a `fix' as follows.  At random sites in the lattice
we introduce a Polyakov--like straight line of magnetic current of
corrective charge $-Q_\mu$ for each direction, and use these as
sources for a dual vector potential. Such lines represent static
monopoles and a random gas of these can lead to a string tension. This
introduces a systematic error to the monopole string tension that we
need to estimate. We do so by placing the same corrective loop on an
otherwise empty lattice, along with a second loop of charge $+Q_\mu$
at another random site. From this new ensemble we calculate the string
tension from Creutz ratios. One half of this is a crude estimate of
the bias introduced in correcting the original configurations, and
this is quoted as a second error on our string tension values, as
appropriate.

\subsection{The largest cluster}
\label{ssec_largest}

\begin{figure}[p]
\begin{center}

\leavevmode

\hbox{
\epsfxsize = 3in
\epsffile{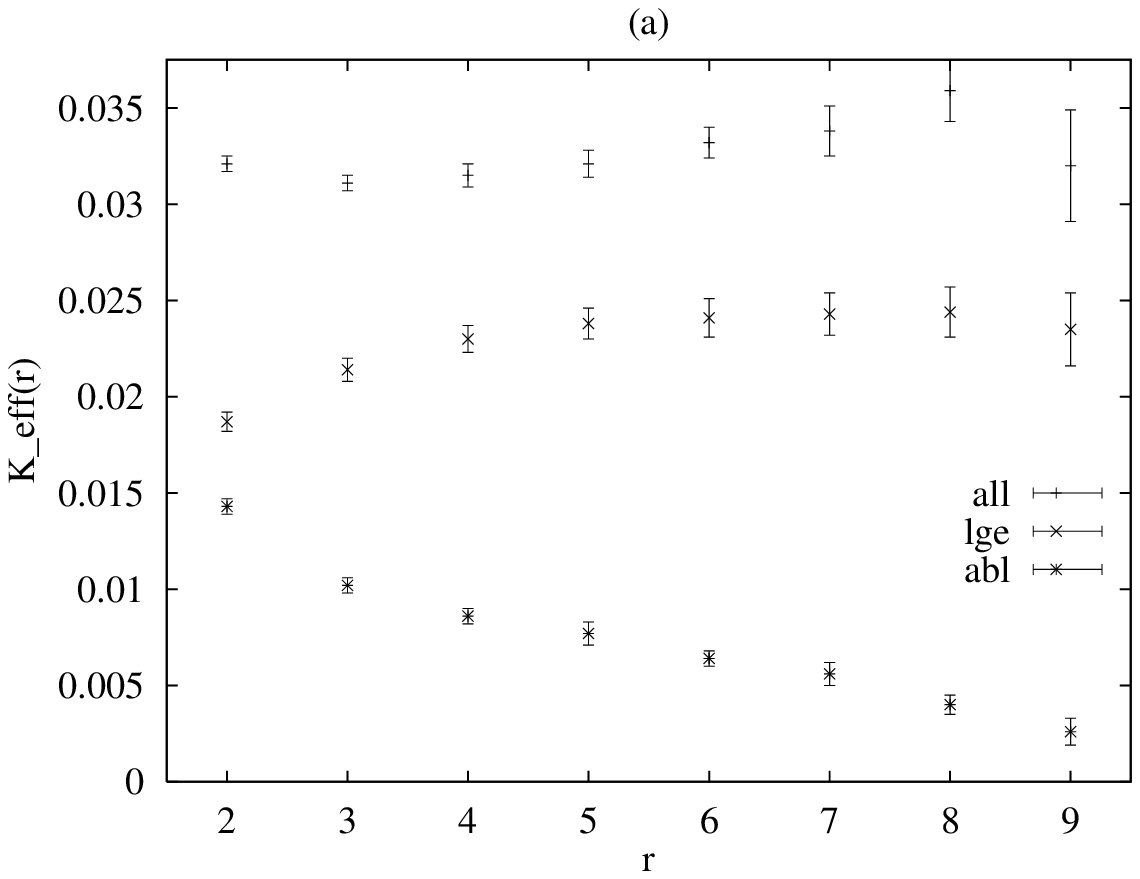}
\epsfxsize = 3in
\epsffile{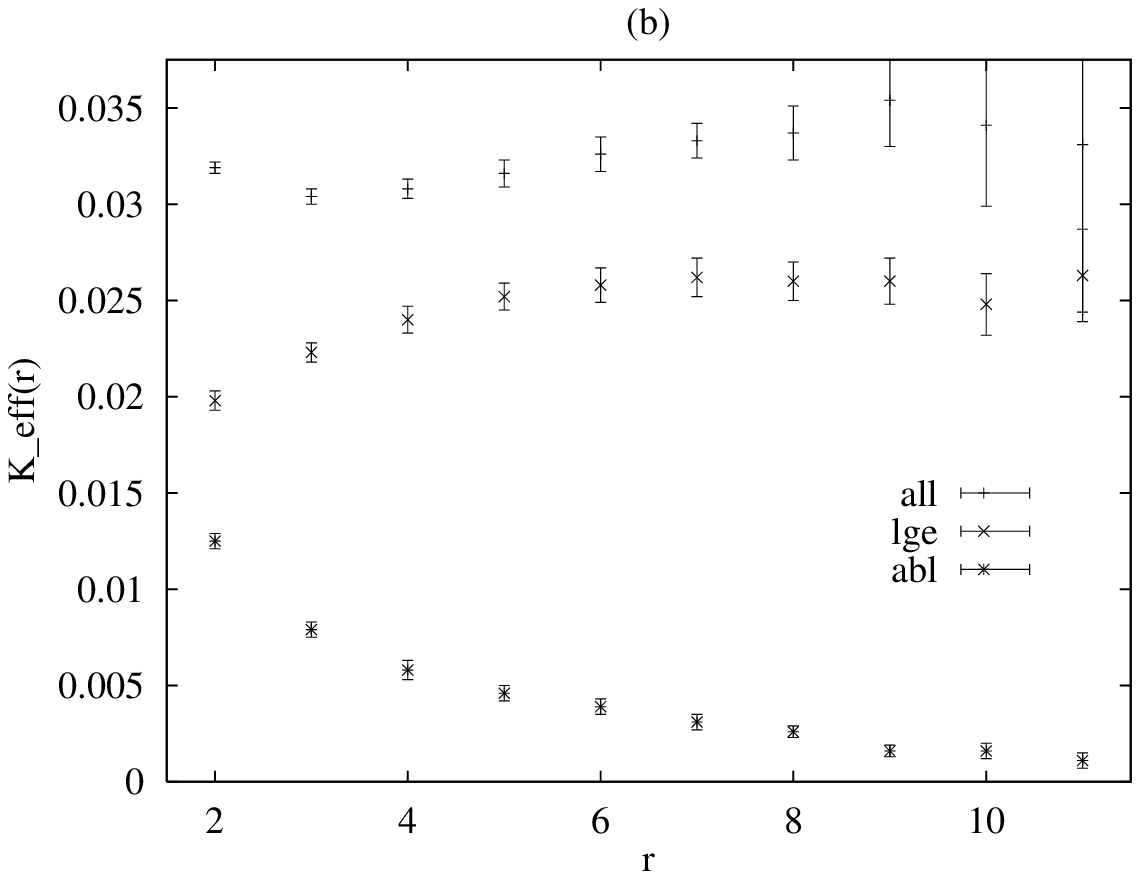}
}

\hbox{
\epsfxsize = 3in
\epsffile{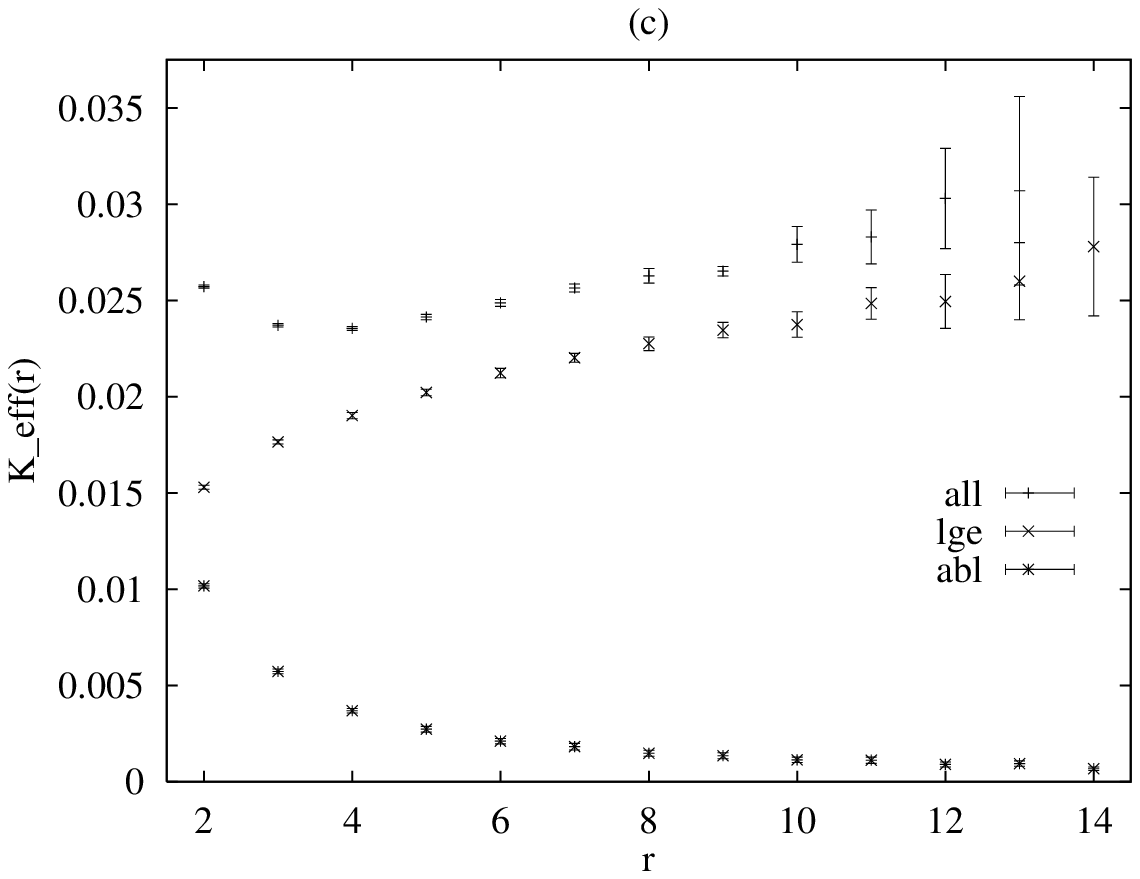}
}
\end{center}
\caption{The effective string tension for all clusters (`all'), the
largest cluster alone (`lge') and the remaining clusters (`abl') on
lattices a) $\beta = 2.5$, $L = 16$, b) $\beta = 2.5$, $L = 20$ and
c) $\beta = 2.5115$, $L = 32$.}
\label{fig_cr_eff}
\end{figure}

In
\cite{hart98}
we observed that at $\beta=2.3$, $L \ge 10$ and at $\beta=2.4$, $L \ge
14$ the $q=1$ monopole string tension was produced almost entirely by
the largest cluster, and the other clusters had a string tension near
zero.  At $\beta=2.5$, $L=16$ the situation was more confused; the
smaller, power law clusters still had a very low string tension, but
that of the largest cluster alone was substantially less than the full
monopole string tension. This suggested some kind of constructive
correlation between the two sets of clusters.  In our new calculation
on an $L=20$ lattice at $\beta=2.5$, we still find a situation that is
confused, although somewhat less so than on the $L=16$ lattice while
on the $L=32$ lattice at $\beta = 2.5115$ the clear picture seen at
$\beta=2.3$ re-emerges, with nearly all the string tension being
produced by the largest cluster, and the remaining clusters having a
negligible contribution. To illustrate this we display in
Figure~\ref{fig_cr_eff} the effective string tensions as a function of
$r$ for the lattices at $\beta=2.5$ and $\beta=2.5115$.

The confused {\it r\^{o}les} of the clusters on finer lattices
\cite{hart98}
is thus a finite volume effect and does not represent a breakdown of
the monopole picture as we near the continuum limit.  Due to the
differing scaling relations for the lengths of the two largest
clusters, it is not enough to maintain a constant lattice volume in
physical units to reproduce the physics as we reduce the lattice
spacing. Rather the lattice must actually become larger even in
physical units, as discussed in subsection~\ref{ssec_thiscalc}.

The string tension arises from `disordering' --- {\it i.e.} switches
in sign --- of the Wilson loop by the monopoles. A monopole that is
sufficiently close to a large Wilson loop will multiply the loop by
$\exp [iq\pi]$ which would na\"{\i}vely suggest that even--charged
loops are not disordered and have no string tension. In a screened
monopole plasma, however, as the monopole is moved away from the loop,
the flux falls and the possibility for disorder and a string tension
exists. [This will also occur without screening, but only when the
monopole is a distance away from the Wilson loop that is comparable to
the size of the loop.] Clearly the exact value of the string tension
will depend upon the details of the screening mechanism, especially as
we increase $q$. This can be calculated in the usual saddle point
approximation
\cite{polyakov77}
where one finds that the string tension is proportional to $q$
\cite{ambjorn98}.
One can obtain a crude model estimate with much less effort,
and this we do in the next subsection. Returning to our
lattice calculations, we list in Table~\ref{tab_sig_mon} 
the monopole string tensions that we obtain using 
charge $q$ Wilson loops at $\beta=2.5115$ on the $L=32$ lattice.
We see that they are indeed consistent with a scaling relation 
$R(q) = q$, at least up to $q=4$.

\subsection{A simple model}
\label{ssec_simp_mod}

It is useful to consider here a simple model for the disordering of
Abelian Wilson loops of various charges by monopoles. We consider only
static monopoles in $d=4$, with a mean field type of screening,
assuming that the macroscopic, exponential fall--off in the flux with
screening length $\xi$ could be applied on the microscopic scale
also. For numerical reasons we also impose a cut--off: beyond 
$N$ screening lengths the flux is set exactly zero. The magnetic
field is thus
\be
B(d) = \left\{
\begin{array}{ll}
\frac{1}{2d^2}e^{-d/\xi} & d \le N \xi \\
0 & d > N \xi.
\end{array}
\right.
\ee
The flux from a monopole distance $z \le N\xi$ above a large
(spacelike) Wilson loop through that loop is
\be
\Phi(N,z,\xi) = \pi \int^1_{z/N\xi} dy \cdot \exp 
\left( - \frac{z}{y\xi} \right).
\ee
Considering a slab of monopoles and antimonopoles all distance $z$
above the loop (and similar below), the charge $q$ Wilson loop gives a
string tension
\cite{hart98}
\be
\delta K(N,z,q) \propto \left( 1 - \cos \left[
q \Phi(N,z,\xi) \right] \right).
\ee
Integrating over all $|z| \le N\xi$, the ratio of string tensions
calculated using charge $q$ and charge $q=1$ Wilson loops is
\be
R(N,q) \equiv \frac{K(N,q)}{K(N,q=1)} = 
\frac
{\int_0^N da \left( 1 - \cos \left[ q\pi\int^1_{a/N} dy \cdot e^{-a/y}
\right] \right)}
{\int_0^N da \left( 1 - \cos \left[ \pi\int^1_{a/N} dy \cdot e^{-a/y}
\right] \right)}
\ee
for this static monopole assumption. This may be evaluated
numerically, and extrapolated as $N \to \infty$, where there is a
well--defined limit, $R(q)$. The results for small $q$ are shown in
Table~\ref{tab_sig_rat}, where the error on $R(q)$ reflects the
extrapolation uncertainty. Comparing these numbers to the actual ratio
of string tensions, we find this simplistic model works remarkably
well for $q=2$, but becomes less reliable as we increase $q$. This
no doubt reflects the increasing importance of the neglected
fluctuations of the flux away from the mean screened values.

\subsection{Monopoles or vortices?}
\label{sec_mon_or_vor}

The fact that the Abelian fields that one extracts in the
maximally Abelian gauge, and their corresponding monopoles,
successfully reproduce the SU(2) fundamental string tension,
provides some evidence for the dual superconducter model
of confinement. As we remarked in the introduction, however,
an attractive alternative picture exists, based on vortex
condensation, and one has comparable evidence for that
picture, obtained by going to maximal centre gauge
and calculating Wilson loops using the singular vortices
obtained after centre projection.

Since the Abelian projected fields seem to contain the
full string tension, it is reasonable to assume that
they encode all the significant confining fluctuations
in the SU(2) fields, even if these are vortices.
How would one expect a vortex to be encoded in the
Abelian fields? And how can we test for their presence?   

Recall that the kind of vortex we are interested in
has a smooth core and flips the sign of any Wilson
loop that it threads. Consider now a space-like
Wilson loop in some time-slice of our Abelian projected
lattice field. We observe that it will flip its sign if 
threaded by a loop of magnetic flux whose core contains a
total flux equal to $\pi$. If the core size is
not arbitrarily large, so that a 
(large enough) Wilson loop has negligible probability
to overlap with the actual core, then a condensate of such
fluxes will lead to linear confinement. Since the original
SU(2) vortex has a smooth core, the simplest expectation
is that this flux, if it reflects the vortex, should not
have a singular monopole source; rather it should 
be a closed loop of magnetic flux. If its length is 
much larger than the size of the Wilson loop, it
can easily thread the loop an odd number of times and
can disorder it. So the natural way for a Z(2) vortex
to be encoded in the Abelian projected fields is
as a closed loop of magnetic flux, in roughly the same position,
and with a smooth core of roughly the same size. If this is so,
and if vortices are present in the SU(2) fields, we would
expect that our Abelian fields contain two kind of
confining fluctuations; singular magnetic monopoles 
and smooth closed loops of ($\pi$ units of) magnetic flux.
Since these closed loops of flux are smooth they will be hard
to identify individually in the midst of the magnetic fluxes 
generated by the monopoles. Their presence can however
be easily tested for as follows. The flux in the U(1) fields 
is conserved and so any flux either originates on the monopoles 
or closes on itself as part of a closed flux loop. The monopoles
are easy to identify and their flux can be calculated.
So for any Wilson loop, $\cal{C}$, we can calculate the 
flux, $B_{\mon}(\cal{C})$, due to the monopoles and
we can subtract it from the total flux, $B[\cal{C}]$,
so as to obtain the remaining flux,
\be
B_{\delta}({\cal{C}}) \equiv B({\cal{C}}) - B_{\mon}({\cal{C}}), 
\label{eqn_flux_diff}
\ee
that comes from closed flux loops. The corresponding value of the
Wilson loop will be $e^{-B_{\delta}(\cal{C})}$.
In this way we can calculate the potential
due to the non-monopole flux, and if we find a
non-zero string tension this demonstrates the existence of
a condensate of such flux loops and provides evidence for 
corresponding Z(2) vortices. If the flux loops carry $\pi$ units 
of flux, Wilson loops corresponding to sources with an even 
charge will have zero string tension.  

We remark here that in U(1) lattice gauge theories, such loops of
magnetic flux are not usually discussed as significant degrees of
freedom. That is not because they cannot exist but rather that the
dynamics is such that they usually play no significant {\it
r\^{o}le}. [One can always smoothly reduce the usual U(1) action by
increasing the core size of such a loop. Ultimately they contribute a
non-confining `spin wave' contribution to the interaction.] The
Abelian projected fields, on the other hand, are not generated from
some local U(1) action. They may possess any structures that are
kinematically allowed.

Vortices can also be encoded in the Abelian fields in a more subtle
way than the above. This involves long-distance correlations amongst
the monopoles.  In $d=2+1$ suppose that at least some of the monopoles
lie along `lines' in such a way that each monopole is followed by an
antimonopole (and vice versa) as we follow the line.  This will
generate an alternating flux of $\pm\pi$ along the line
\cite{deldebbio98}. 
So a Wilson loop threaded by this line will acquire a factor of
$-1$. Such correlated ensembles can therefore encode the vortices in
the original three dimensional SU(2) fields. A similar restriction of
monopole current world lines to two dimensional sheets can be
envisaged in $d=3+1$. In both cases, their presence would be signalled
by the fact that they do not disorder Wilson loops corresponding to an
even charge (unlike a plasma of monopoles). So if we calculate the
string tension due to the monopoles, and if we find a significant
suppression of the $q=2$ string tension, then this will indicate the
significant presence of such correlations and hence of vortices.
 
This latter way of encoding vortices in the Abelian
projected fields might seem less natural given the
smoothness of the underlying Z(2) vortices. As pointed out in
\cite{ambjorn98},
however, such correlated monopole structures actually occur in what
one usually regards as a standard example of a field theory that
demonstrates linear confinement driven by monopole condensation: the
Georgi-Glashow model in three dimensions. This model couples an SU(2)
gauge field to a scalar Higgs field in the adjoint representation of
the gauge group. The theory has a Higgs phase, and the Higgs field
drives the gauge field into a vacuum state which has only U(1) gauge
symmetry, save in the cores of extended topological objects. These 't
Hooft--Polyakov monopoles are magnetically charged with respect to the
U(1) fields, and give rise to the linear confining potential, at least
in the semi--classical approximation
\cite{polyakov77}
which holds good when the charged vector bosons are heavy.
As pointed out in
\cite{ambjorn98},
however, this conventional picture cannot be true on large enough
length scales since eventually the presence of the charged massive
$W^{\pm}$ fields will lead to the breaking of strings between doubly
charged sources (the $W^{\pm}$ possessing twice the fundamental unit
of charge).  A plasma of monopoles, on the other hand, will predict
the linear confinement of such double charges. So it was argued that
in this limit it is Z(2) vortices, which do not disorder doubly
charged Wilson loops, that drive the confinement
\cite{ambjorn98}.
The crossover between the two pictures, it is argued, would occur
beyond a certain length scale dictated by the $W^{\pm}$ mass, where
the distribution of monopole flux would no longer be purely Coulombic,
but would be collimated into structures of lower dimension ---
essentially strings of alternating monopoles and antimonopoles ---
that reflect the Z(2) vortices of the vacuum.

Of course one cannot carry this argument over in all its details
to the case of the pure SU(2) gauge theory. Here there are no
explicit Higgs or $W^{\pm}$ fields; any analogous objects would need
to be composite. The theory also has only one scale, and so
one would not expect an extended intermediate region between 
the onset of confining behaviour and the collimation of the flux
signalling Z(2) disorder. But  it does raise the
possibility that the Z(2) vortices in the SU(2) fields might
be encoded, after Abelian projection, in such correlations amongst 
the monopoles rather than in separate smooth closed loops of magnetic 
flux.  

To probe for the presence of smooth loops of magnetic flux
in the Abelian projected fields, we have calculated the 
`difference' flux, as defined in eqn.~\ref{eqn_flux_diff}, 
and the resulting string tension; and to probe for 
vortex-like ensembles of monopoles we have calculated
the monopole string tension, $K(q)$, for various source 
charges, $q$. 

We start with the latter. In Table~\ref{tab_sig_mon_eff}
we show the $q=1,2$ monopole effective string tensions that
we have obtained from Creutz ratios on the  $L=32$ lattice
at $\beta = 2.5115$. We see that for $q=2$, just as 
for $q=1$, there are very few transients at small $r$, and
the extraction of an asymptotic string tension appears
to be unambiguous.  We have accurate calculations out to a 
distance of $r=9a$ which corresponds to
\be
r=9a \sim \frac{1.6}{\sqrt{K}}
\label{eqn_dist_K}
\ee
in physical units at this $\beta$. Out to this distance
there is absolutely no hint of any reduction in the 
$q=2$ effective string tension. It has been pointed out
\cite{deldebbio98}
that when the Wilson loop is not much larger than the typical vortex
core, it is not completely unnatural to obtain an effective string
tension comparable to the one from a monopole plasma.  Here the size
is beginning to be large compared to the natural scale of the theory,
however, and it is hard not to view the lack of any variation at all
in the $q=2$ effective string tension as pointing to the absence of
the kind of correlations amongst the monopoles that might be encoding
Z(2) vortices.

The second possibility is that the vortices might be
encoded not in correlations amongst the monopoles but rather
in closed loops carrying $\pi$ units of magnetic flux. 
Such loops would contribute to $K(q=1)$ but not to $K(q=2)$. 
We recall that there has been a calculation of $K(q)$,
calculated within the full Abelian fields at 
$\beta = 2.5115$, and that there it was found
\cite{bali96}
that there is a finite $q=2$  effective string
tension that extends out to at least as far as $r=9a$, and 
that the ratio of the U(1) string tensions is 
$K(q=2)/K(q=1) = 2.23$~(5). While this suggests that
closed flux loops are not important, these
string tensions necessarily include the contribution from
monopoles, and it would be useful to have a calculation
that excludes the latter.
We have therefore calculated the effective
string tension using only the flux that comes from closed flux
loops, as defined in eqn.~\ref{eqn_flux_diff}. The
results of this calculation are listed 
in Table~\ref{tab_sig_mon_eff} for $q=1$. We see that, within
small errors, there is
asymptotically no string tension from such loops (a
potential fit to the Wilson loops yields $K < 0.0025$). 
This shows in a direct way that there is no significant condensate
of closed loops of flux in the Abelian projected fields.

In conclusion, our investigations here have shown no
sign of vortices encoded in the Abelian projected fields in 
either of the two ways that one might plausibly have
expected them to be.

It is worth stepping back at this point and reflecting 
upon the tentative nature of the above arguments.
Our calculation of the monopole string tension takes
each monopole to be a source of a simple Coulombic flux, 
as obtained by solving Maxwell's equations. Treating
the monopoles as being `isolated' in this way, is
the obvious starting point if one wishes to ask what 
is the physics `due to monopoles'. But it is no guarantee 
that such a question makes any sense. Indeed it is only
in the Villain model that one has the exact factorisation of
Wilson loops into monopole and non-monopole pieces that
is needed for this question to be clearly unambiguous.
For example, it is not {\it a priori} clear that the ensemble 
of monopoles one obtains in the maximally Abelian gauge 
is even qualitatively such as one would
expect from a generic U(1) action. If it is not, then one
must ask what are the fluctuations in the SU(2) fields that
determine the nature of the monopole ensemble, and whether
these features of the ensemble have a significant effect on
the calculated string tension. If they do then the question 
we are asking, whether the string tension is `due to monopoles', 
becomes intrinsically ambiguous. Our demonstration
that there is no suppression of the $q=2$ monopole string tension
may be regarded as a first step, but only a first step,
towards showing that the monopole ensemble does 
not possess such features that require additional explanation. 
One should also mention that the Abelian fields are
periodic in $2\pi$ (in the sense that the number density
of plaquette angles peaks at multiples of $2\pi$), which is 
the requirement for Dirac strings to be invisible, and that they
possess a screening length that is characteristic of plasmas
\cite{hart98}.
Equally, if we had found a significant flux loop condensate in
the Abelian fields, we would have had to study 
carefully the (presumably) non-trivial correlations between
the monopoles and flux loops in order to determine if there 
was any sense in claiming that some physics was `due to
monopoles'. The fact that we have not found any sign
of such a flux loop condensate, or of any anomalous
features of the monopole plasma, means that we are not
yet forced to confront this quite general problem. But this
question clearly needs systematic exploration.

\section{Summary}
\label{sec_summ}

We have studied the magnetic monopole currents obtained after fixing
to the maximally Abelian gauge of SU(2), on lattices that are both
large in physical units and have a relatively small lattice
spacing. The monopole clusters are found to divide into two clear
classes both on the basis of their lengths and their physical
properties. The smaller clusters have a distribution of lengths which
follows a power law, and the exponent is consistent with 3, as was
previously seen on coarser lattices
\cite{hart98}.
These clusters are compact objects, and their radii also follow a
power law whose exponent we found to be 4.2~(1). This is close to, but
a little less than, the scale invariant value of 5 which indicates
that if the smaller clusters correspond to objects in the SU(2)
vacuum, these objects have a size distribution which yields slightly
more large radius objects than would be expected in a purely scale
invariant theory. This scale breaking is, however, far too weak to
encourage the identification of such objects with the small instantons
in the theory.

That is not to say that instantons are necessarily irrelevant; the
correlations between the monopole currents and the action and
topological charge densities (
\cite{feurstein97,bakker98,hart96,inst99}
and references therein)
indicate some connexion. It would be interesting to measure
the correlations separately using the largest cluster, and the
remaining, power law clusters. 

The small clusters do not appear relevant to the long range physics;
they produce a zero, or at most very small, contribution 
to the string tension.
Indeed the string tension is consistent with being produced by the 
largest cluster alone. The fact that there should be a large
percolating monopole cluster associated with the long-distance
physics is an old idea (see
\cite{polikarpov93}
for an early reference). The properties that we find for this cluster,
however, are certainly not those associated with na\"{\i}ve
percolation. In particular, as we approach the continuum limit the
density of monopoles belonging to this cluster goes to zero. And
indeed the fraction of the total monopole current that arises from
this largest cluster also appears to go to zero. This is because this
single very large cluster seems to percolate on physical and not on
lattice length scales, while the physically unimportant secondary
clusters have an approximatley constant density in lattice units.  All
this reproduces the properties that we previously obtained on coarser
lattices, but which seemed to be lost when going to finer lattice
spacings, albeit on volumes of a smaller physical extent. This study
demonstrates that the breakdown was a finite volume effect, rather
than a failure of the monopole picture in the weak coupling limit. The
volume at which the picture was restored was as predicted by the
scaling relations derived from the coarser lattices.

The fact that one has to go to volumes that are ever larger
as $a \to 0$, can be interpreted as a breakdown of
the Abelian projection. As we remarked, something like this
is not unexpected: as $a\to 0$ the Abelian projection will
presumably be increasingly driven by the irrelevant
ultraviolet fluctuations of the SU(2) link matrices.
This leads to an increasing fraction of the monopole
current -- that belonging to the smaller clusters --
containing no physics and this contributes an increasing 
background `noise' to attempts at extracting physical 
observables as we approach the continuum limit. Fortunately
the unphysical gas of monopoles that one obtains by
Abelian projection within the maximally Abelian gauge
is sufficiently dilute that one can isolate the 
physically relevant `percolating' cluster, even if the
price is that one has to work with ever larger volumes.

We also calculated the monopole contribution to
Wilson loops of higher charges, and found
that the corresponding monopole string tensions  
appear to be simply proportional to the charge,
at least up to $q=4$. This is what is predicted by a
saddle point treatment of the U(1) lattice gauge theory
\cite{ambjorn98}
as can be seen more simply, if more approximately,
within our simplistic charge plasma model.

Our main reason for studying these higher-$q$ string tensions, was to
probe for any sign of a condensate of Z(2) vortices in the Abelian
projected fields. It might, of course, be that such vortices are
simply not encoded in the Abelian fields. It is plausible, however, to
infer from the observed monopole and centre dominance that both when
we force the SU(2) link matrices to be as Abelian as possible, and
when we force them to be as close to $\pm 1$ as possible, the
resulting Abelian and Z(2) fields capture essentially all the long
range confining disorder present in the original SU(2) fields. In the
case of Z(2) fields the disorder must be encoded by vortices (there
is nothing else).  In the Abelian case however the disorder can be
carried either by monopoles or by closed loops of `magnetic' flux.  We
argued that such a closed loop, carrying a net magnetic flux of $\pi$
units, provides a plausible way for the Abelian fields to encode the
presence of an underlying Z(2) vortex.  Our study of the
monopole--U(1) `difference gas' showed, however, that there is no
significant contribution to confinement from such loops of magnetic
flux. An alternative
\cite{deldebbio98,ambjorn98}.
is that the Z(2) disorder is encoded in correlated
strings of (anti)monopoles. If such correlations
were important, however, they would lead to
a significant suppression of the $q=2$ string
tension, and this  we do not observe. Instead we find that
the effective monopole string tensions satisfies
$K(q=2) = 2 K(q=1)$ very accurately to distances that  
are quite substantial in physical units. While there is a 
limit to what one can conclude about Z(2) vortices in
a study that focuses solely on the Abelian projected fields, 
the fact that they do not manifest
themselves in any of the ways that one might expect,
must cast some doubt on their importance in the SU(2) vacuum.

\subsection*{Acknowledgments}

The gauge fixed $L=32$ field configurations were crucial to the 
work of this paper and we are very grateful to Gunnar Bali for 
making them available to us. 
Our computations were performed on a UKQCD workstation and 
this work was supported in part by United Kingdom PPARC grant
GR/L22744. The work of M.T. was supported in part by PPARC grant
GR/K55752.

\begin{table}[p]
\begin{center}
\begin{tabular}{llr*{4}{r@{ $-$ }l}}
\hline \hline
\multicolumn{1}{c}{$L$} & 
\multicolumn{1}{c}{$\beta$} & 
\multicolumn{1}{c}{$N$} &
\multicolumn{2}{c}{$l_{\max}$} & 
\multicolumn{2}{c}{$l_{\second}$} & 
$l_{\max}$ & $l_{\second}$ & 
\multicolumn{2}{c}{$l_{\third}$} \\ 
\hline
 12 &  2.3    & 500 & 2358 & 3970   & 18 & 220   & 2172 & 3930              \\
 14 &  2.4    & 500 & 894 & 3436    & 22 & 1112  & 28 & 3400                \\
 16 &  2.5    & 500 & 268 & 2462    & 22 & 910   & 4 & 2414                 \\ 
 20 &  2.5    & 100 & 1718 & 5050   & 50 & 1644  & 318 & 4964   & 36 & 684  \\
 32 &  2.5115 & 100 & 11872 & 20040 & 114 & 4676 & 9066 & 19886 & 92 & 2476 \\ 
\hline \hline 
\end{tabular}
\caption{Range of lengths found for the
largest, the second largest, their difference and for third largest 
clusters for the
ensembles of $N$ configurations shown.}
\label{table_length} 
\end{center}
\end{table}

\begin{table}[p]
\begin{center}
\begin{tabular}{lllll}
\hline \hline
& $q$ & \multicolumn{1}{c}{$K$ --- all} & \multicolumn{1}{c}{$K$ --- lge} & 
\multicolumn{1}{c}{$K$ --- abl} \\
\hline
$\beta = 2.5$, $L=20$
& 1 & 0.035 (3) & 0.026 (3) (1) & $< 0.0015$ \\
\hline
$\beta = 2.5115$, $L=32$
& 1 & 0.0270 (10) & 0.0240 (10) (3) & $< 0.0010$ \\
& 2 & 0.0520 (10) & 0.0450 (10) (4) & $< 0.0022$ \\
& 3 & 0.075 (2) \\
& 4 & 0.103 (5) \\
\hline \hline
\end{tabular}
\end{center}
\caption{Monopole string tensions from Wilson loops
of varying charge, using all current (`all'), current from the
largest cluster alone (`lge') and the remaining current (`abl').}
\label{tab_sig_mon}
\end{table}

\begin{table}[p]
\begin{center}
\begin{tabular}{lll}
\hline \hline
$q$ & \multicolumn{1}{c}{$R(q)$} & \multicolumn{1}{c}{$K(q)/K(q=1)$} \\
\hline
2 & 1.827 (1) & 2.00 (9) \\
3 & 2.192 (1) & 2.88 (14) \\
4 & 2.526 (1) & 3.96 (25) \\
\hline \hline
\end{tabular}
\end{center}
\caption{Ratio of monopole string tensions from Wilson loops
of varying charge, in the static plasma model and measured
at $\beta = 2.5115$ on $L=32$.}
\label{tab_sig_rat}
\end{table}

\begin{table}[p]
\begin{center}
\begin{tabular}{llll}
\hline \hline
$r$ & 
\multicolumn{1}{c}{$K_{\eff}(r)$, $q=1$} & 
\multicolumn{1}{c}{$K_{\eff}(r)$, $q=2$} & 
\multicolumn{1}{c}{$K_{\eff}^{\diff}(r)$, $q=1$} \\
\hline
2 & 0.02572 (8)  & 0.0511 (2)   & 0.0799 (2) \\
3 & 0.02371 (9)  & 0.0497 (2)   & 0.0335 (2) \\
4 & 0.02355 (10) & 0.0483 (3)   & 0.0176 (5) \\
5 & 0.02414 (14) & 0.0492 (5)   & 0.0102 (9) \\
6 & 0.02487 (17) & 0.0496 (7)   & 0.0082 (13) \\
7 & 0.02565 (20) & 0.0501 (7)   & 0.0062 (21) \\
8 & 0.02628 (37) & 0.0493 (25)  & 0.0076 (53) \\
9 & 0.02652 (25) & 0.0556 (58)  & -0.0076 (27) \\
\hline
\end{tabular}
\end{center}
\caption{Effective monopole string tensions from Creutz ratios for 
charges $q=1,2$ 
and from the difference of U(1) and monopole
fluxes at $\beta = 2.5115$ on $L=32$.}
\label{tab_sig_mon_eff}
\end{table}

\begin{thebibliography}{99}

\bibitem{hart98}
A. Hart and M. Teper,
Phys. Rev. D 58 (1998) 014504, 
available as hep-lat/9712003. 

\bibitem{mandelstam76}
S. Mandelstam,
Phys. Rep. 23 (1976) 245.

\bibitem{thooft81}
G. 't Hooft,
Nucl. Phys. B 190 (1981) 455.

\bibitem{kronfeld87}
A.S. Kronfeld, M.L. Laursen, G. Schierholz and U.J. Wiese,
Phys. Lett. B 198 (1987) 516.

\bibitem{suzuki90}
T. Suzuki and I. Yotsuyanagi,
Phys. Rev. D 42 (1990) 4257.

\bibitem{stack94}
J.D. Stack, S. Nieman and R.J. Wensley,
Phys. Rev. D 50 (1994) 3399,
available as hep-lat/9404014.

\bibitem{bali96}
G.S. Bali, V. Bornyakov, M. M\"uller--Preussker and K. Schilling,
Phys. Rev. D 54 (1996) 2863,
available as hep-lat/9603012.

\bibitem{feurstein97}
M. Fuerstein, H. Markum and S. Thurner,
Phys. Lett. B 398 (1997) 203;
E.-M. Ilgenfritz, H. Markum, M. M\"{u}ller--Preussker, W. Sakuler 
and S. Thurner,
Progr. Theor. Phys. Suppl. 131 (1998) 353,
available as hep-lat/9804031.

\bibitem{bakker98}
B.L.G. Bakker, M.N. Chernodub, M.I. Polikarpov and A.I. Veselov,
hep-lat/9811001.

\bibitem{thooft79}
G. 't Hooft,
Nucl. Phys. B 138 (1978) 1; B 153 (1979) 141.

\bibitem{mack80}
G. Mack,
in `Recent Developments in Gauge Theories', ed. G. 't Hooft et al.
(Plenum, New York, 1980).

\bibitem{nielsen79}
H. Nielsen and P. Olesen,
Nucl. Phys. B 160 (1979) 380.

\bibitem{kovacs98}
T. Kov\'acs and E.T. Tomboulis,
Phys. Rev. D 57 (1998) 4054, 
available as hep-lat/9711009.

\bibitem{deldebbio98}
L. Del Debbio, M. Faber, J. Greensite and \v{S}. Olejn\'{\i}k,
Phys. Rev. D 58 (1998) 094501,
available as hep-lat/9802003.
M. Faber, J. Greensite and \v{S}. Olejn\'{\i}k,
Phys. Rev. D 57 (1998) 2603,
available as hep-lat/9710039.

\bibitem{ambjorn98}
J. Ambj{\o}rn and J. Greensite,
J. High Energy Phys. 9805 (1998) 004,
available as hep-lat/9804022.

\bibitem{polyakov77}
A. Polyakov,
Nucl. Phys. B 120 (1977) 429;
Gauge Fields and Strings (Harwood Academic, Chur, Switzerland, 1987).

\bibitem{degrand80}
T. DeGrand and D. Toussaint,
Phys. Rev. D 22 (1980) 2478.

\bibitem{hart97a}
A. Hart and M. Teper,
Phys. Rev. D 55 (1997) 3756,
available as hep-lat/9606007. 

\bibitem{hioki91}
S. Hioki, S. Kitahara, Y. Matsubara, O. Miyamura, S. Ohno and
T. Suzuki,
Phys. Lett. B 271 (1991) 201.

\bibitem{hart97b}
A. Hart and M. Teper,
Nucl. Phys. B (Proc. Suppl.) 53 (1997) 497,
available as hep-lat/9606022. 

\bibitem{hart96}
A. Hart and M. Teper,
Phys. Lett. B 371 (1996) 261, 
available as hep-lat/9511016.
V. Bornyakov and G. Schierholz,
Phys. Lett. B 384 (1996) 190, 
available as hep-lat/9605019.
M. Fukushima et al,
Phys. Lett. B 399 (1997) 141, 
available as hep-lat/9608084.

\bibitem{inst99}
S. Sasaki and O. Miyamura, 
Phys. Rev. D 59 (1999) 094507,
available as hep-lat/9811029.
M. Fukushima, H. Suganuma and H. Toki, 
hep-lat/9902005.

\bibitem{grady98}
M. Grady, 
Phys. Lett. B 455 (1999) 239,
available as hep-lat/9802035.

\bibitem{stack92}
J.D. Stack and R.J. Wensley,
Nucl. Phys. B 371 (1992) 597.

\bibitem{polikarpov93}
T. Ivanenko, A. Pochinskii and M. Polikarpov,
Phys. Lett. B 302 (1993) 458. 
 
\end{thebibliography}
\end{document}